\documentclass[%
 reprint, 
 longbibliography,
floatfix,
 amsmath,amssymb,
aip,
jcp,
 superscriptaddress
]{revtex4-1}

\usepackage[english]{babel}

\usepackage{graphicx}

\usepackage{xcolor}
\definecolor{negg}{RGB}{238,238,239}
\definecolor{nblue}{RGB}{61,180,229}

\usepackage[hyperfootnotes=false]{hyperref}	 
\hypersetup{
  colorlinks,
  citecolor=nblue,
  linkcolor=nblue,
  urlcolor=nblue
}

   \usepackage[scaled=0.95]{helvet}
 \usepackage{tgtermes}

\usepackage[nameinlink]{cleveref}

 \crefname{equation}{{\color{black}{Eq.}}}{{\color{black}{Eqs.}}}
 \creflabelformat{equation}{#2\textup{(#1)}#3}
 \crefname{figure}{Fig.}{Figs.}
 \Crefname{figure}{Figs.}{Figs.} 
 \crefname{table}{Table}{Tables}
 \crefname{appendix}{{\color{black}{Sec.}}}{{\color{black}{Secs.}}}
 \crefname{section}{{\color{black}{Sec.}}}{{\color{black}{Secs.}}}

 \usepackage{caption}
\usepackage{ragged2e}

 \DeclareCaptionJustification{justified}{\justifying}
\newcommand{\del}{\partial}
\newcommand{\delfrac}[2]{\frac{\del #1}{\del #2}}

\newcommand{\tb}[1]{_{\text{\tiny #1}}}
\newcommand{\tu}[1]{^{\text{\tiny #1}}}

\newcommand{\eps}{\varepsilon}
\newcommand{\kB}{k_\text{B}}
\newcommand{\kT}{\kB T}
\newcommand{\vpull}{v\tb{pull}}

\newcommand{\pan}[1]{{\bfseries #1}}

\begin{document}

\title{Nonequilibrium friction and free energy estimates for kinetic coarse-graining - Driven particles in responsive media}


%

\author{Sebastian Milster}%
\affiliation{Physikalisches Institut, Albert-Ludwigs-Universität Freiburg, 
D-79104 Freiburg, Germany}

\author{Joachim Dzubiella}
\affiliation{Physikalisches Institut, Albert-Ludwigs-Universität Freiburg, D-79104 Freiburg, Germany}
\affiliation{Cluster of Excellence livMatS @ FIT – Freiburg Center for Interactive Materials and Bioinspired Technologies, Albert-Ludwigs-Universität Freiburg, D-79110 Freiburg, Germany}

\author{Gerhard Stock}
\affiliation{Physikalisches Institut, Albert-Ludwigs-Universität Freiburg,
D-79104 Freiburg, Germany}

\author{Steffen Wolf}%
\email{steffen.wolf@physik.uni-freiburg.de}
\affiliation{Physikalisches Institut, Albert-Ludwigs-Universität Freiburg,
D-79104 Freiburg, Germany}

\begin{abstract}
Predicting the molecular friction and energy landscapes under nonequilibrium conditions is key to coarse-graining the dynamics of selective solute transport through complex, fluctuating and responsive media, e.g., polymeric materials such as hydrogels, cellular membranes or ion channels. The analysis of equilibrium
ensembles already allows such a coarse-graining for very mild nonequilibrium conditions. Yet in the presence of stronger external driving and/or inhomogeneous setups, the transport process is governed apart from a potential of mean force also by a nontrivial position- and velocity-dependent friction. It is therefore important to find suitable and efficient methods to estimate the mean force and the friction landscape, which then can be used in a low-dimensional, coarse-grained Langevin framework to predict the system's transport properties and timescales. In this work, we evaluate different coarse-graining approaches based on constant-velocity constraint simulations for generating such estimates using two model systems, which are a 1D responsive barrier as a minimalistic model and a single tracer driven through a 3D bead-spring polymer membrane as a more sophisticated problem. Finally, we demonstrate that the estimates from 3D constant-velocity simulations yield the correct velocity-dependent friction, which can be directly utilized for coarse-grained (1D) Langevin simulations with constant external driving forces.

\end{abstract}

\maketitle

\section{Introduction}


\begin{figure*}
    \centering
    \includegraphics[width=0.9\textwidth]{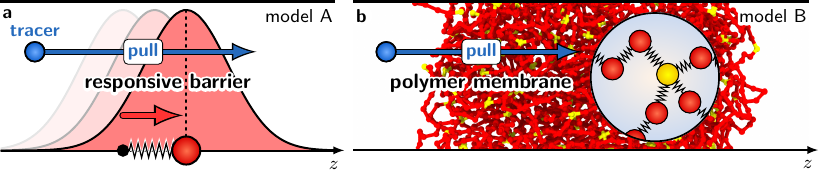}
    \caption{{\bfseries a}: Sketch of the minimal model A. The tracer (blue bead) is pulled along the 1D coordinate $z$ across a Gaussian barrier. The dynamic barrier position represented by a point particle (sketched as red bead) is attached to a fixpoint (black bullet) with a harmonic spring, furthermore follows fluctuating/dissipative dynamics, and is responsive to the (Gaussian) tracer--barrier interactions. When the tracer is pulled across the barrier, the barrier gets also pushed to the right, and eventually snaps/slips back, leading to both elastic and dissipative interactions. {\bfseries b}: Snapshot of the 3D polymer membrane model B. The polymer is a bead-spring network (see circular inset), with red chain monomers and yellow cross-linkers. The tracer (blue bead) is pulled from the solvent domain through the fluctuating membrane, and interacts with the polymer via a Lennard-Jones potential.}
    \label{fig:intro_figure}
\end{figure*}


Molecular transport is of fundamental importance for numerous processes and applications involving functional soft matter, ranging from biological phenomena to medical and industrial applications, such as cellular homeostasis,\cite{Keener1998Oct} protein--ligand binding,\cite{Copeland2006,Pan2013} particle separation,\cite{Geise2010Aug,Agboola2021Feb,Falk2015Apr} tissue engineering,\cite{Brohem2011Feb,Wichterle1960Jan} drug delivery,\cite{Moncho-Jorda2020Nov,Stamatialis2008Feb,Hoffman1987Dec} and nanocatalysis.\cite{Lu2011Jun,Angioletti-Uberti2015Jul,Roa2017Sep} Typically, nonequilibrium conditions are the rule rather than the exception for transport phenomena. In fact, a significant number of processes of interest includes external driving, such as the transport of molecules or particles within nanostructures, soft matter and biophysical systems as well as materials.\cite{diVentra08,Leitner09,Loos2024Feb,Liese2017Jan,Best10,Schulz12,Hu16,Rico19} Prime examples include ion diffusion through membrane channels driven by an electrostatic gradient, \cite{Berneche01,Dzubiella2005Jun,Maffeo2012Dec,Kopfer14,Jaeger22} proton gradients driving ATP production,\cite{Mitchell61} ion transport through polymeric electrolyte membranes,\cite{Zhang2012May} or self-propelled particles exploring polymer matrices.\cite{Kim2022AugActive}

The tremendous details of nonequilibrium processes in soft matter often require complex models with unattainable computational expenses. Hence, one relies on adequate coarse-grained (CG) descriptions to compute the desired physical observables on meaningful length and timescales. Regarding the transport phenomena, transition rates and fluxes are usually the quantities of interest, i.e., the CG model must reproduce the correct particle dynamics on the desired resolution in space and time. To this end, the particle dynamics are projected onto a collective variable of interest\cite{Zwanzig01,Mori65,Grabert82} describing the `system'. The remaining orthogonal degrees of freedom are integrated out and render the `bath'. This strategy has been proven to correctly identify the relevant microscopic interactions\cite{Wolf20} and predict the effective dynamics of the system.\cite{Straub87,Singer1994Sep,Schulz12,Setny13,Post22,Dalton24,Milster24} 

Determining equilibrium quantities is usually well understood,\cite{Chipot} and the results directly enter Langevin or Fokker-Planck equations.\cite{Risken1996} However, nonequilibrium scenarios are far more challenging and only local theories and approaches have been derived.\cite{Zwanzig01,Oettinger05,Haenggi97,Fortini2014Oct} Notable challenges include the appearance of velocity-dependent friction,\cite{Shinjo1993Mar,Erdmann2000May,Buchholtz1998May,Gutierrez-Varela2021Sep,Takehara2010Dec, Zhu1990Sep,Takehara2014Apr,Voigtmann2013Nov,Fusco2005Jan,Loos2024May,Matsukawa1994Jun} the influence of memory\cite{Meyer19,Meyer20,Milster24,Schmid2023Feb,Ariskina2024Sep,Klippenstein2021May, Dalton2024Oct,Straube2020Jul,Jung2017Jun,Loos2024May} due to an insufficient timescale separation between system and bath dynamics,\cite{Wolf18,Post22,Schilling2022Aug} and a limited validity of the fluctuation-dissipation theorem.\cite{Koch24}

For a generalized consideration of such transport processes, we here focus on a single tracer particle that is driven by an external force $f\tb{ext}$ through a complex environment with constant number of particles $N$ and volume $V$ along a coordinate $z$. The force $f\tb{ext}$ may represent a constant force, e.g., from driving by an electrostatic potential, or a time- and/or position-dependent force $f\tb{ext}(z,t)$. The complete setup is in contact with a thermal bath (solvent) with temperature $T$. The tracer dynamics is subject to the interactions with the surrounding particles, which are coarse-grained into
time-independent quantities, i.e., a potential of mean force (PMF) $F$ and friction $\gamma$, allowing for a description based on the Langevin equation
\begin{align}
      m \dot v &= -\delfrac{F}{z} +  f\tb{ext} - m\gamma v  + \sqrt{2\kT m \gamma} ~\xi ,\label{eq:CGLangevin}
\end{align}
with Gaussian white noise $\xi(t)$ ($\langle\xi(t)\rangle=0$, $\langle\xi(t)\xi(t')\rangle=\delta(t-t')$, and thermal energy $\kT=\beta^{-1}$. In general, the complex environment is not homogeneous in space, and hence not only $F$, but also $\gamma$ is position-dependent.

Furthermore, we are particularly interested in scenarios with strong driving, where the dynamics differ significantly from equilibrium or mild nonequilibrium cases. Hence, the dissipation is expected to depend on the velocity $v$, i.e.,
\begin{align}
\gamma\to\gamma(z,v).  \label{eq:gamma_z_v} 
\end{align}
In general, extracting the PMF $F(z)$ and the position- and velocity-dependent friction landscape can be challenging. Determining $F$ from simulations is a well-understood problem, for which a variety of methods exist.\cite{Chipot,Kosztin2006Feb} In equilibrium, the friction coefficient $\gamma(z,v)$ can be directly related to the diffusion coefficient via the second fluctuation-dissipation theorem as $D=\kT/(m \gamma\tb{EQ})$\cite{Zwanzig01} and can be computed using, e.g., the mean squared displacement, velocity autocorrelations, or the memory kernel.\cite{Kubo1966Jan,BarratHansen,Straube2020Jul} While these methods may already require good sampling and a careful analysis, the situation becomes even more demanding in nonequilibrium ($f\tb{ext}\neq0$) beyond linear response, particularly due to $\gamma(z,v)\neq\gamma\tb{EQ}$. To this end, the Dudko-Hummer-Szabo model\cite{Dudko06} or dissipation-corrected targeted molecular dynamics\cite{Wolf18} (dcTMD) provide efficient estimates for the energy and friction landscapes, including the velocity-dependent dynamics. 

Noteworthy, the sampling of rare events is optimized when external driving is implemented as a constant-velocity constraint with velocity $\vpull$, for which the constraint force $f_{\rm c}(z)$ is calculated via a Lagrangian multiplier. This serves as the basis for the estimation of both PMF and friction. Furthermore, setting $f\tb{ext}=f_{\rm c}(z)$ and taking the ensemble average ($\langle\dots\rangle$) \cref{{eq:CGLangevin}} simplifies to\cite{Wolf18}
\begin{equation}
                 -\delfrac{ F(z)}{z}  + \langle f\tb{ext}(z)\rangle - m\gamma(z,\vpull) \vpull = 0,  \label{eq:meanfield}
\end{equation}
allowing for convenient analysis.

In this work, we compare the applicability of different coarse-graining approaches using two model systems with increasing complexity (see \cref{fig:intro_figure}) for different strengths of external driving and coarse-grain both systems along the pulling dimension $z$. 
The numerically inexpensive model A (\cref{fig:intro_figure}a) consists of a tracer particle pulled over a responsive barrier, while in model B (\cref{fig:intro_figure}b), the tracer is driven through a molecular representation of a model membrane. In both cases, the full microscopic dynamics are projected onto $z$ as ''system'' collective variable, while the remaining degrees of freedom (barrier dynamics in model A, membrane particles in model B) constitute the ''bath'' degrees of freedom. Model A allows for a reliable comparison between the estimators, and furthermore reveals local velocity-dependent bath dynamics contributing to the friction. Model B is more complex and closer to a relevant system for real applications such as a hydrogel. 


Finally, we also perform constant-force simulations ($f\tb{ext}$ = const.) for model B and demonstrate that the coarse-grained $F(z)$ and $\gamma(z,v)$ estimates can be directly used for 1D Langevin simulations (\cref{eq:CGLangevin}). The resulting coarse-grained mean particle velocity for passage through the model system agrees well with the velocity observed in corresponding constant-force simulations with model B.

\section{Theory}
Assuming that pulling is described by Eq.~(\ref{eq:CGLangevin}), we introduce four different ways to estimate the PMF $F$ as well as the friction $\gamma$ in the following. Note that we set $F(0)=0$ and therefore use the symbol $F(z)$ instead of the commonly used $\Delta F(z)$ for convenience.
Our analysis procedure assumes a constant pulling velocity $\vpull$ along the reaction coordinate $z$, and estimates $F(z)$ and $\gamma(z,v)$. A further requirement is a clear timescale separation between the pulling and all other degrees of freedom, i.e., the surrounding bath medium still exhibits equilibrium dynamics despite the non-equilibrium pulling. This usually holds for small $\vpull$. However, we will see that the following estimators are in principle also suitable for higher velocities, provided that the bath particles are in equilibrium before the interaction/collision and sufficient sample trajectories are available.

\subsection{Decomposition of the total work and Jarzynski's equality}
In general, the work performed on the tracer during the pulling in each trajectory, 
\begin{align}
 W(z) = \int_0^z\mathrm{d}z' f\tb{ext}(z'),
\end{align}
can be split into $F(z)$, i.e., the reversible work, and the dissipation contribution $\langle W\tb{diss}(z)\rangle$, i.e., the irreversible work as 
\begin{equation}
 \langle W (z)\rangle = F (z) + \langle W\tb{diss}(z)\rangle \label{eq:workcontributions}
\end{equation}
with the brackets $ \langle ... \rangle $ denoting an ensemble mean over a set of statistically independent trajectories starting from a common Boltzmann state distribution.
The dissipation is directly related to the friction coefficient when pulled with a constant velocity constraint within the framework of the Markovian Langevin equation [\cref{eq:CGLangevin}], reading\cite{Wolf18}
\begin{align}
    \gamma(z,\vpull)=\frac{1}{m\vpull}\delfrac{\langle W\tb{diss}(z)\rangle}{z}.\label{eq:friction_general}
\end{align}



In this work we employ two work-based estimates for $F(z)$ and $\langle W\tb{diss}(z)\rangle$, which both are based on Jarzynski's equation\cite{Jarzynski97} 
\begin{align}
     F\tb{Jarz}(z) &= - \kT  \log \left< e^ {-\beta W(z) } \right>.
    \label{eq:Jarzy}
\end{align}


Since $\langle W(z)\rangle$ is known from simulations,~\cref{eq:workcontributions,eq:friction_general,eq:Jarzy} provide the corresponding friction
\begin{align}
    \gamma\tb{Jarz}(z,\vpull)=\frac{1}{m\vpull}\delfrac{ }{z}\left(\langle W(z)\rangle-F\tb{Jarz}(z)\right). \label{eq:gammaJarz}
\end{align}

\subsection{Dissipation-corrected targeted MD (dcTMD)}

Since \cref{eq:Jarzy} converges very slowly,\cite{Wolf20} dcTMD estimate uses a 2\textsuperscript{nd}-order cumulant expansion of  \cref{eq:Jarzy}, 
which is exact if $W(z)$ is normally distributed. The PMF based on dcTMD thus reads
\begin{align}
F_W(z) = \langle W(z)\rangle-\frac{ \beta}{2} \left< \delta W(z)^2 \right>.
\end{align}
with the fluctuation $\delta W = W - \left< W \right>$.
The corresponding friction [\cref{eq:friction_general}] can then be calculated via the variance of the work,\cite{Wolf18} reading
\begin{align}
    \gamma_W(z,\vpull) &= \frac{\beta}{2 m\, \vpull} \delfrac{ }{z} \left< \delta W(z)^2 \right>.\label{eq:gammaW}
\end{align}

We demonstrated earlier\cite{Wolf23} that the validity of the 2\textsuperscript{nd}-order approximation may highly depend on $\vpull$, e.g., if various process pathways exist. An indication for this effect is an overestimation of $\gamma$ and thus an underestimation of $F$.



%

\subsection{PMF-based friction}\label{sec:macroscopic}
Suppose $F(z)$ is known from other sources, we define the \emph{PMF-based friction} as 
\begin{align}
    \gamma\tb{PMF}(z,\vpull)=\frac{1}{m \vpull}\left(\langle f\tb{ext}(z) \rangle-\delfrac{F(z)}{z}\right), \label{eq:macroscopic_friction}
\end{align}
which directly follows from \cref{eq:meanfield} and is in line with \cref{eq:friction_general}, since we identify $\langle W\tb{diss}(z)\rangle=\langle W(z)\rangle -F(z)$. Conveniently, it only depends on the local ensemble averages of the forces.
\cref{eq:macroscopic_friction} follows the general concept that the frictional force must balance all other forces in the steady-state, which is well-established\cite{Risken1996} and investigated in more detail with simulations with constant-force pulling.\cite{Gazuz2013Mar,Gnann2011Feb,Gazuz2009Jun} To the best of our knowledge, and despite the simplicity of \cref{eq:macroscopic_friction}, this constant-velocity interpretation has not been reported yet. Notably, it yields the advantage that it directly provides the velocity-dependence, particularly in the high-velocity regime.

\subsection{Boxed PMF and friction profiles}\label{sec:bipartide}
In some cases, geometric considerations of a system indicate the general shape of $F(z)$ and $\gamma(z,v)$. 
In the case of model B, we may distinguish between the symmetric central membrane region and the surrounding (implicit) solvent. We therefore define a smooth box-shaped function
\begin{align}
\phi(z)= \exp\left[-\left(\frac{z-z\tb{ com}}{2d}\right)^q\  \right], \label{eq:tildephi}
\end{align}
and will use the adjective \emph{boxed} in this work relating to this approach. In \cref{eq:tildephi}, the are three free parameters: the effective width $d$ of the membrane region, the membrane's center of mass $z\tb{com}$, and an even integer  $q$ controlling the shape, i.e., the steepness/width of the interfaces (e.g., Gaussian: $q=2$,  discrete domains: $q\to\infty$). For very short interaction ranges, $\phi(z)$ is essentially proportional to the polymer volume fraction. 

We assume a linear relation between $F(z)$, $\gamma(z,v)$ and $\phi(z)$, hence, also in the boxed form as
\begin{align}
F\tb{box}(z)&=F\tb{in}\ \phi(z)\label{eq:Gz}\\
\gamma\tb{box}(z,v)&=\gamma\tb{0}+(\gamma\tb{in}(v)-\gamma\tb{0})\ \phi(z),\label{eq:GAMMAz}
\end{align}
with the values $F\tb{in}$ and $\gamma\tb{in}$ inside the membrane. We note that non-linear scalings of mean values of $F$ and $\gamma$ with $\phi$ may be found for polymers.\cite{Yasuda1969,Amsden1998,Milster24,Kim2020}  

In the case of constant velocity pulling, we can utilize the mean force profiles $\langle f\tb{ext}(z)\rangle$ to fit $z\tb{com}$, $d$, $q$, $F\tb{in}$ based on \cref{eq:meanfield,eq:tildephi,eq:Gz,eq:GAMMAz}, using the PMF-based friction \cref{eq:macroscopic_friction} averaged inside the membrane for $\gamma\tb{in}(v)$. The above analytical expressions are particularly beneficial for CG Langevin simulations. Note that the functional form of the velocity-dependence $\gamma\tb{in}(v)$ will be provided in \cref{eq:gamma_fit} in the Results section.


\section{Methods\label{sec:microscopic}}

\subsection{Model A: responsive-barrier crossing }
    
Model A is a minimal model to study the velocity-dependent friction of driven particles through a responsive environment. Precisely, a single particle, the ''tracer'', is pulled across a Gaussian barrier, which may respond to the tracer's presence by confining the barrier position to a harmonic potential instead of fixing it. 
It is modeled by two stochastic particles (one particle representing the tracer, one the barrier) along one dimension $z$ and a constant environment friction $\gamma\tb{0}$. Its form bears semblance to the Prandtl–Tomlinson model,\cite{Prandtl1928Jan,Tomlison1929Jun,Fusco2005Jan,Dong2011Dec} in which a Brownian particle, attached to a moving spring, interacts with a sinusoidal potential. Although the Prandtl-Tomlinson model can be treated to some extent analytically,\cite{Muser2011Sep,Gnecco2012Jul} our model provides insights for a single barrier crossing, which can be interpreted as a (soft) collision of the tracer with a single bath particle of similar mass.

The tracer dynamics are described by a Langevin equation. The external driving $f\tb{ext}$ can be chosen such that it ensures that $v=\vpull$ and $m\dot v = 0$, and hence
\begin{align}
0=&-\delfrac{U\tb{int}(z-z\tb{b})}{z}  +f\tb{ext}(z,t)\nonumber\\
&-m \gamma\tb{0} \vpull+ \sqrt{2\kT m \gamma\tb{0}}\ \xi(t), \label{eq:tracer_eom}    
\end{align}
where $U\tb{int}$ is the interaction potential between the tracer and the barrier (bath) particle. The dynamics of the barrier described by $z\tb{b}$ and $ v\tb{b}$ read
\begin{align}
m\dot{v}\tb{b}=&-\delfrac{U\tb{int}(z\tb{b}-z)}{z\tb{b}}-\delfrac{U\tb{spr}(z\tb{b})}{z\tb{b}}\nonumber\\
&-m \gamma\tb{0} v\tb{b} + \sqrt{2\kT m \gamma\tb{0}}\ \xi\tb{b}(t),\label{eq:barrier_eom}   
\end{align}
with $U\tb{int}$ the response to the tracer. The barrier is also attached to the system's center by a harmonic spring $U\tb{spr}$. The tracer and the barrier are subject to dissipation and fluctuations from the solvent with a friction constant $\gamma\tb{0}$.

The interaction and the spring potential read
\begin{align}
U\tb{int}(y)&=\varepsilon\tb{b}\exp\left[-2\frac{y^2}{\sigma\tb{b}^2}\right]\\
U\tb{spr}(y)&=\frac{k}{2}\left(y-L/2\right)^2    
\end{align}
with $\varepsilon\tb{b}$ the (maximum) interaction energy, $\sigma\tb{b}$ the barrier width (interaction radius), $k$ the spring constant, and $L$ the system size. The spring constant corresponds to different degrees of responsiveness to the tracer and thus fluctuations of the barrier. 
While the tracer particle moves along $z$ with constant speed, the barrier mid point will first be pushed by the tracer to larger values of $z$ and move back towards its center point when $U\tb{int}(z(t)) < U\tb{spr}(z(t))$.
Small $k$ yields high responsiveness with large fluctuations, while high $k$ result in a rather stiff barrier with low responsiveness and little fluctuations. In this work, we will fix the barrier height to $\beta\varepsilon\tb{b}=3$, and probe different spring constants $k/k\tb{0}$ and pulling velocities $\vpull/v\tb{0}$, with $k\tb{0}=\kT/\sigma\tb{b}^2$ and $v\tb{0}=\sigma\tb{b}\gamma\tb{0}$ the respective reference values.
Further simulation details are provided in the Appendix \labelcref{app:seim_model_A}.

\subsection{Model B: polymer membrane permeation }\label{sec:modelB}

Model B represents a slab-shaped polymer membrane (see \cref{fig:intro_figure} b) that is periodic in the $x$-$y$ directions, finite in $z$, and is located in the center of the simulation box with box lengths $L_x=L_y\approx12.2\sigma$, $L_z\approx42.7\sigma$, and bead size $\sigma$. The polymer is organized in the form of a molecular sieve consisting of $1920$ chain monomers (red) that are connected by $104$ cross-linker beads (yellow). Monomers as well as beads are connected via harmonic bond and angle potentials, locally forming a tetrahedral topology. All particles interact via (nonbonded) Lennard-Jones (LJ) potentials. The force field is identical to our previous work \cite{Milster24} with parameters yielding a volume fraction of roughly $32\%$.
The tracer is pulled across the membrane in $z$-direction starting in the pure solvent region. Note that we apply constant-velocity and constant-force pulling simulations. Hence, the general dynamics of the tracer (with index $i=1$) and the polymer beads ($i \neq 1$) are governed by the Langevin equation
\begin{align}
m\dot{\mathbf{v}}_i(t) = & \mathbf{f}\tu{int}_i(t) + f_{i, \text{\tiny ext}}(t)\mathbf{e}_z\nonumber\\
&-\gamma\tb{0} m\mathbf{v}_i(t) + \sqrt{2\kT m \gamma\tb{0}}\ \boldsymbol{\xi}_i(t) ,\label{eq:CLE}  
\end{align}
where $\mathbf{f}\tu{int}_i(t)$ accounts for all interparticle interactions and $f_{i, \text{\tiny ext}}$ a biasing force, which acts only on the tracer and only along $z$.  

The LJ potential between the tracer and any of the membrane particles reads
\begin{align}
    U(r)=4\varepsilon\left[ \left(\frac{\sigma}{r}\right)^{12} -\left(\frac{\sigma}{r}\right)^{6} \right],
\end{align}
with distance $r$, and interaction strength $\varepsilon$. For the latter, we chose three different values, $\beta\varepsilon\in\left\{0.1, 1.0, 2.0\right\}$, ranging from purely repulsive to highly attractive interactions, and probe different pulling velocities $\vpull/v\tb{0}$ (with reference $v\tb{0}=\sigma\gamma\tb{0}$). See Appendix \labelcref{app:seim_model_B} for further simulation details.




\subsection{Implementation of biases for constant velocity\label{sec:v_constraint}}
For both models A and B, we start driving at $z_0$ where the tracer interaction with the remaining degrees of freedom is negligible (see again Fig.~\ref{fig:intro_figure}), and we can here set $F(z_0)=0$. For the sake of simplicity, we use the following constant velocity implementation: First, we perform an unbiased integration step to obtain the total force $f\tb{unbiased}$ acting on the tracer in $z$-direction. Subsequently, we shift the tracer along the $z$-coordinate to fulfill the velocity constraint $v=\vpull$. Note that the constraint does not act on the $x$-$y$-direction (in the case of the 3D model B). In such an integration step including the biasing shift, the external driving $f\tb{ext}$ is supposed to compensate the unbiased forces (since $m\dot v=0$), and hence, we assume $f\tb{ext}=-f\tb{unbiased}$. 

Note that in dcTMD the constant velocity is usually achieved by a constraint in the form of a Lagrange equation of motion of the first kind,\cite{Wolf18,Post22} i.e., the constraint force performs work on the tracer (and its environment) at the very time step. In our simple biasing approach, on the contrary, the work is performed indirectly through the subsequent unbiased integration step. Interestingly, we will see that this is suitable for the prediction of friction and energy profiles, assumingly because the two constraint methods converge for sufficiently small time increments.


Moreover, for convenience and less noisy data, we take advantage of the fact that the underlying solvent friction and fluctuations are additive, and that the motion along $z$ is already defined by $\vpull$. Precisely, we collect only tracer--barrier (or tracer--polymer) forces, i.e., neglecting $-m\gamma\tb{0} \vpull$ and $\sqrt{2\kT m \gamma\tb{0}}\xi(t)$ [see \cref{eq:tracer_eom,eq:barrier_eom}]. After the analysis one may simply add the known solvent effects  ($\gamma\tb{0}$) to the final results.


\section{Results and Discussion}
\subsection{Model A: responsive barrier crossing}
\subsubsection{Friction and energy profiles\label{sec:modelAprofiles}}

\begin{figure}[tb]
\includegraphics[width=1\linewidth]{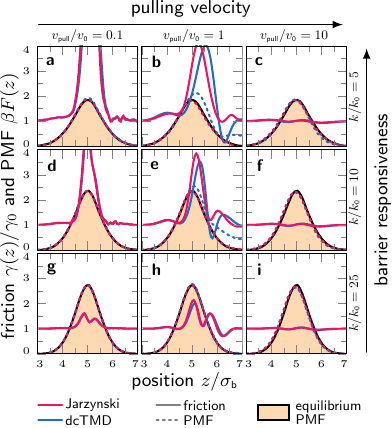}
\caption{Examples of the dissipation-correction analysis for the responsive barrier (model A) depicting the position dependent friction and PMF for different (barrier) spring constants $k/k\tb{0}=\in\{5,10,25\}$ (compare row labels an on the r.h.s.) and pulling velocities $v\tb{pull}/v\tb{0}\in\{0.1,1,10\}$ (column labels). The dashed lines depict the PMF and the friction profile is presented by solid lines, while the method is further color-coded (legend below panels). 
The shaded area is the PMF obtained from equilibrium distributions ($p(z)$) via $\beta F\propto \-\log(p(z))$. Note that the PMF-based friction is not displayed, as it is practically identical with the Jarzynski estimate.}
\label{fig:1D_domains}
\end{figure} 

In \cref{fig:1D_domains} we display the PMF and friction estimates for the 1D fluctuating barrier model for equilibrium as well as for different pulling velocities (column labels above panels) and barrier responsiveness (row labels on the r.h.s.). 
The equilibrium PMFs exhibit lower and broader profiles with decreasing $k/k\tb{0}$, since the barrier spreads along $z$ due to thermal fluctuations.
The PMFs for intermediate and large $k/k\tb{0}$ in turn are narrower but larger owing to a smaller fluctuation range.

Let us now consider the PMF and friciton profiles from non-equilibrium pulling in dependence of barrier responsiveness: 
For all tested pulling velocities and the responsiveness range, the Jarzynski-based PMF estimates excellently match the equilibrium profiles. We note that this agreement is owed to the large number ($N=10000$) of independent simulations we can produce for this small test system, which circumvents the notoriously slow convergence of the estimator.\cite{Wolf20} The cumulant-based estimator mostly matches the equilibrium PMF except for the intermediate velocity range of $v\tb{pull}/v\tb{0}=1.0$ and the highly as well as the moderately responsive barrier ($k/k\tb{0}$=5 and 10, respectively) for reasons that we discuss below.

Concerning the friction estimates (solid lines), we first consider estimates at the small pulling velocity $v\tb{pull}/v\tb{0}=0.1$ (see \cref{fig:1D_domains}a,d,g), where the pulling timescale is typically slower than the barrier's response time. The friction estimates are highest with the smallest $k/k\tb{0}$ and centered around the midpoint of $z$. In a corresponding microscopic interpretation, the driven tracer applies work on the barrier by displacing it from its equilibrium position. As soon as $U\tb{int}(z(t)) < U\tb{spr}(z(t))$, the barrier accelerates back towards its center point, and its dynamics irreversibly remove the previously applied work, providing the dissipation channel for the system. A large responsiveness of the barrier leads to a larger variance of $f_{\rm ext}$ within the trajectory ensemble and consequently a larger $W\tb{diss}$.
With increasing $k/k\tb{0}$, i.e., decreasing responsiveness, the spread of barrier mid-point positions around their center point is decreasing, resulting in a smaller force as well as work variance within the traejctory ensemble, leading to a smaller friction.


For a pulling velocity of $v\tb{pull}/v\tb{0}=1$ (see \cref{fig:1D_domains}b,e,h), we find mixed system/bath timescales.
 Of particular interest is a deviation between the two PMF estimations in this velocity regime. The Jarzynski-based estimator reproduces the correct PMF for all $k$, and hence the correct friction profiles. The cumulant-based estimate increasingly deviates from the equilibrium estimate with the responsiveness, resulting in an overestimated PMF for large $z$.
 The friction coefficients in turn become skewed to higher values of $z$ (\cref{fig:1D_domains}b, $k/k\tb{0}=5$), and can even exhibit oscillations (see \cref{fig:1D_domains}h, $k/k\tb{0}=25$) that is present for both estimators. The latter observation is already evidence of some imperfect system-bath timescale separation, which we relate to the intricate tracer--energy transfer during the collision (see Supplementary Movies 1--3): for the highly responsive barriers ($k/k\tb{0}=5$, Supplementary Movie 1), we observe splitting of the barrier ensemble into barriers before and after successful jump, which represents an emergence of `dynamic subpopulations' corresponding to pathways in the bath degree of freedom [\cref{eq:barrier_eom}], causing a deviation of the work distribution $p(W(z))$ from a normal distribution. From earlier investigations, we know that such a pathway formation leads to a wrong friction and thus incorrect PMF estimate by dcTMD.\cite{Wolf23} For weakly responsive barriers ($k/k\tb{0}=25$, Supplementary Movie 3), the barriers oscillate once after transition and apply a net force to the particle, which appears as a minimum in friction in both Jarzynski- and cumulant estimations.




Lastly, the situation at fast pulling $v/v\tb{0}=10$ (see \cref{fig:1D_domains}c,f,i) is markedly different from the other cases. Concerning the PMF, we observe profiles that are identical to the ones estimated at slow pulling. However, the friction is practically identical with $\gamma\tb{0}$ over the whole range of $z$. This observation can be explained by a system-bath reversal: the tracer moves so fast that the barrier has no time to respond, and the latter effectively becomes stationary. The tracer therefore experiences no fluctuations, i.e., no friction, except the predefined $\gamma\tb{0}$ from the solvent. The situation is reminiscent of the superlubric regime described by Shinjo and Hirano,\cite{Shinjo1993Mar} and has been observed by us for other molecular systems such as lubricants.\cite{Post22}

\subsubsection{Velocity-dependent friction}

\begin{figure}[tb]
    \centering
    \includegraphics[scale=1.3]{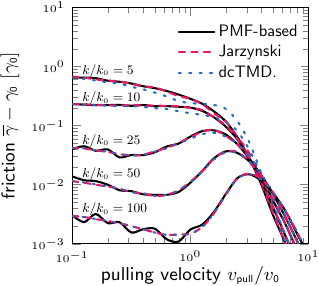}
    \caption{Overall mean friction without solvent contribution (expressed as $L^{-1}\int_0^L \mathrm{d}z\,\gamma(z) - \gamma\tb{0}$) of model A vs. the pulling velocity $\vpull$ shown for different barrier spring constants $k$. 
    \label{fig:frictionModelA}}
\end{figure}


Having identified the connection between PMF and friction estimates within the three dominant pulling velocity regimes, we now investigate how the system develops from one of those regimes to another. \cref{fig:frictionModelA} displays the relative friction contribution from the fluctuating barrier, expressed as $L^{-1}\int_0^L\mathrm{d}z \gamma(z)-\gamma\tb{0}$, which depends on responsiveness as well as the pulling velocity.
Despite the simplicity of the system, we observe a rich behavior in friction changes.  In general, the friction increases with higher barrier responsiveness due to the increased fluctuations and the ability to absorb energy, and at very high $\vpull$, the friction vanishes for all tested $k$, attributed to the tracer becoming faster than the bath degree of freedom.
The most interesting feature are the friction minimum and maximum in the range of $0.1<\vpull/v\tb{0}<5$ and $k/k\tb{0} > 10$, resulting from barrier resonance-like effects. The friction maximum occurs at half of the damped harmonic oscillator resonance frequency rescaled by the barrier width  $\vpull\tu{max}=\frac{\sigma\tb{b}}{\pi}\sqrt{k/m-\gamma\tb{0}^2/4}$. Microscopically speaking, the barrier is accelerated back after the tracer traversed the peak, and oscillates back and forth. During one oscillation, the tracer has already left the possible interaction region, and thus no work is returned to the tracer. Conversely, minimization occurs at a slower velocity, roughly at $\vpull\tu{min}=\frac{1}{4}\vpull\tu{max}$, at which the barrier catches up with the tracer during one oscillation, pushing it from smaller $z$ and thus returning some of its previously absorbed energy. Both effects are most pronounced for large $k$ since the barrier behaves almost deterministic following the above microscopic dynamics. For small $k/k\tb{0}$, dissipation from the barrier fluctuations dominates and suppresses deterministic effects. Interestingly. we recognize qualitative similarities with the resonant activation known from overdamped barrier crossing,\cite{Schmitt2006May,Doering1992Oct} but, in our case, the responsiveness $k$ and pulling velocity $\vpull$ somehow take the role of the \emph{barrier modulation}.

In general, model A exhibits dynamics resembling the macroscopic effects of shear thinning and thickening. Only in the case of $k/k\tb{0}=10$, we observe a Newtonian behavior, i.e., a velocity-independent friction for $\vpull/v\tb{0}<1$, where the `deterministic' non-monotonic friction and the fluctuation-induced friction supposedly balance.

All these characteristic are recovered by cumulant- and the Jarzynski-based estimators as well as our PMF-based reference, confirming their applicability with sufficient sampling. Only the cumulant approximation deviates for very responsive barriers and for a pulling slightly faster than the `resonance velocity'. This we attribute to the possible emergence of `dynamic subpopulations', explained in \cref{sec:modelAprofiles}, which is further supported by the fact that the cumulant-based estimator provides the correct friction for the rather deterministic barriers at larger $k$.

In summary, in this very simple test model we observe interesting nonequilibrium effects that appear during external driving of a macroscopic system, effectively coarse-graining the resulting nonequilibrium dynamics into a PMF and a velocity-dependent friction $\gamma(v)$. 

    

\subsection{Model B:  polymer membrane permeation }
\subsubsection{Friction and energy profiles}
\begin{figure}[tb]
    \centering

    \includegraphics[width=1\linewidth]{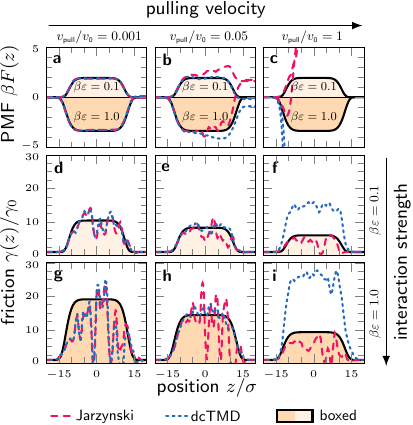}
    \caption{Model B: comparison of different estimates for the free energy (panels \pan{a} to \pan{c}) and  
    friction profiles (panels \pan{d} to \pan{i}) at different pulling velocities $\vpull/v\tb{0}\in\left\{0.001, 0.05, 1\right\}$ (see column labels) and with different tracer--network interaction strengths $\beta\varepsilon\in\left\{0.1, 1.0\right\}$ (see row labels). \label{fig:model_B_profiles} \label{fig:dcTMD_membrane} }  
\end{figure}

Having established the capabilities to coarse-grain the PMF and the dynamics of the 1D test system in the form of a energy and friction profiles, we move to the more challenging system of model B. 
\cref{fig:dcTMD_membrane} displays the PMF (panels a to c) and friction (panels d to i) estimates for different interaction strengths $\beta\varepsilon\in\left\{0.1, 1.0 \right\}$ (as indicated by row panels and inside panels, respectively), and pulling velocities, $\vpull/v\tb{0}\in\left\{0.001, 0.05, 1\right\}$ (column labels). Additionally, we here employ the boxed-profile approach, as the partitioning of the system allows us to fit the equations provided in \cref{sec:bipartide} to the mean force profiles $\langle f\tb{ext}(z) \rangle$ (see Appendix \labelcref{app:MFfit} for more details).

At very slow pulling ($\vpull/v\tb{0}=0.001$, see \cref{fig:dcTMD_membrane}a,d,g), all estimators yield the same PMF. The friction estimations are in good agreement with each other for $\beta\varepsilon=0.1$ (\cref{fig:dcTMD_membrane}a). For $\beta\varepsilon=1$, the friction estimates are noisy due to the slow convergence behavior of the respective estimates\cite{Wolf18} and slightly underestimated compared to the boxed friction. The PMF-based friction (cf. \cref{fig:macrofriction_profile} in Appendix \labelcref{app:MFfit}) exhibits even larger fluctuation for the very slow pulling (mainly due to the $\vpull^{-1}$-scaling), which are not represented in the boxed evaluation by design.   

\begin{figure*}
    \centering  
    \includegraphics[scale=1.3]{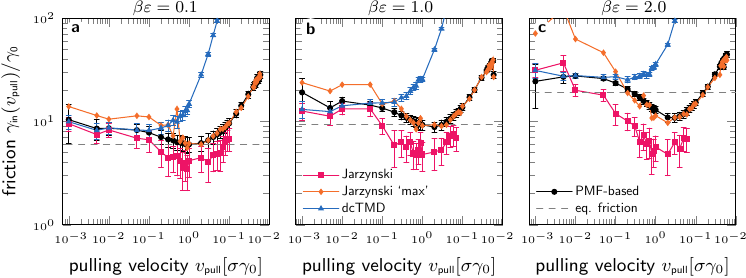}
    \caption{Model B: comparison of different friction estimates inside the membrane vs. the pulling velocity $\vpull$, and for three different interaction strengths $\varepsilon\in\left\{0.1, 1.0, 2.0\right\}$ (see labels above panels). For Jarzynski `max' display the maximum values inside the membrane instead of the mean value, which can  provide good estimates at higher velocities (see main text for full details). The gray dashed line depicts the known equilibrium friction $\gamma_{\rm EQ}$ from unbiased simulations.\cite{Milster24}  \label{fig:gamma_vs_v}}
\end{figure*}

In the case of intermediate pulling ($\vpull/v\tb{0}=0.05$), the Jarzynski and cumulant PMF estimates (\cref{fig:dcTMD_membrane}b) deviate from the expectation, as they do not converge to zero, except the cumulant approximation for weak interactions $\beta\varepsilon=0.1$. Jarzynski's method overestimates the PMF as a consequence of insufficient sampling,\cite{Wolf20} and thus underestimates the friction (\cref{fig:dcTMD_membrane}e,h). The cumulant approximation in turn tends to overestimate the friction and consequently underestimates the PMF, which has been reported before as an effect of interacting system and bath timescales.\cite{Wolf23} 
The box-shaped assumption results in PMF estimates that are in very good agreement with the ones at the small pulling velocity. Its friction estimates are slightly smaller than for the slow velocity, but semi-quantitatively follow the trend of the Jarzynski and the dcTMD estimators.

Finally, for fast pulling ($\vpull/v\tb{0}=1$), the friction (f and i) is overestimated by the dcTDM estimator, and the PMFs (panel c) quickly diverge inside the membrane to very small values.\cite{Wolf23} The Jarzynski-based approach in turn overestimates the PMF and underestimates friction, again most likely due to missing sampling of low-work trajectories. The boxed estimator uses PMF estimates from smaller velocities, depicting even more decreased friciton.

The PMF-based friction, when being supplied with the boxed PMF data, yields well-converged friction estimates for intermediate and high pulling velocities. For small velocities however, these estimates contain more noise than the Jarzynski- and cumulant-based estimates because of limited sampling. 

In summary, the estimates based on Jarzysnki's equality and its 2\textsuperscript{nd} order cumulant approximation perform best under small velocities and interaction strengths. An increase of both parameters leads to the introduction of noise, and even wrong estimates in the case of very fast pulling. The boxed PMF and friction profiles provide consistent results for alle velocities, however, it relies on prior knowledge of the shape / spatial domains. The other estimators are agnostic in this respect and are therefore, in principle, universally applicable. 
Due to their reliability, we therefore use the boxed PMF as reference in the following considerations. To allow for an assessment of the systematic error of the friction estimates, we use the PMF-based friction estimate as reference, in turn.



    

\subsubsection{Velocity-dependent friction inside membrane}

Having tested the spatially resolved performance of different PMF and friction estimators of the membrane model B, we now analyze the friction inside the membrane $\gamma\tb{in}$ with respect to the interaction strength $\beta\varepsilon$ and driving velocity $\vpull$, presented in \cref{fig:gamma_vs_v}.

For slow pulling, we recognize an initial overestimation of $\gamma\tb{in}$ in comparison to the equilibrium friction\cite{Milster24} by a factor of two, which can be related to the constraint affecting the monomer degrees of freedom timescales.\cite{Daldrop17,Wolf18} 
Increasing $v_{\rm pull}$, we first find a decay of $\gamma\tb{in}$ reminiscent of the situation in model A. With increasing velocity, friction minimizes at the typical Lennard-Jones timescale around $\vpull\approx \sigma\gamma\tb{0}$. With increasing interaction strengths, the minimum is shifted to higher velocities due to increased coupling to faster bond vibrations of the polymer with a ''resonance velocity'' of $\vpull\approx\frac{\sigma}{2\pi}\omega\tb{bond}\approx 4\sigma\gamma\tb{0}$ (since $\omega\tb{bond}\approx25$\cite{Milster24}).

In contrast to model A, we observe an eventual friction increase for high velocities $\vpull/v\tb{0} \gtrsim 1$ for all estimators, indicating a real physical effect  reminiscent of shear-thickening. A similar rise in friction has been observed in dcTMD simulations of LJ particles, as well.\cite{Post22} We relate this effect to the ''hard'' repulsion of LJ particles, whereas in model A, the tracer--barrier interactions are too weak to display this effect.
At higher velocities, the hard collisions with the membrane particles become increasingly more likely, and the duration of such a collision becomes shorter, i.e., more kinetic energy is passed to the membrane that cannot be returned to the driven tracer and thus is dissipated. The tracer is also more likely to be scattered orthogonally (in the $x$-$y$--plane),\cite{Voigtmann2013Nov} which is eventually absorbed by the solvent damping. Large driving velocities can even lead to local deformation of the membrane. 

Comparing the fast-pulling behavior for the different estimators, we find an early divergence of the cumulant approximation for $\vpull/v\tb{0} \gtrsim 0.5$, which is connected to a deviation of the work distribution from a bell curve at larger perturbations of the system.\cite{Wolf23} The Jarzynski-based estimate is more stable over a wider range of velocities, but again underestimates the friction at higher velocities compared to the box-shaped reference. We recognize that when only taking the maximal friction estimate along $z$ into account (denoted as `Jarzynski max' in \cref{fig:gamma_vs_v}), we compensate for insufficient sampling at higher velocities, and the Jarzynski estimate agrees well with the boxed friction result. Note that the height of the maximum depends on the width of the bins in which the estimate is calculated. For the presented estimate, a bin width of $\sim$0.7$\sigma$ is apparently ideal for high velocities. 

It is evident from these considerations that $\gamma\tb{in}$ is indeed driving velocity-dependent. To take this dependence into account, we heuristically approximate $\gamma\tb{in}(v)$ as 
\begin{align}
    \gamma\tu{fit}\tb{in}(v)=a_1\exp\left(-a_2 v^{a_5}\right)+a_3 v^{a_4}, \label{eq:gamma_fit}
\end{align}
where the first term controls the initial decay, while the second is responsible for the fast-pulling divergence.
\begin{table}[b]
    \centering
     \caption{Fit parameters of $\gamma\tb{in}(v)$, based on \cref{eq:gamma_fit} and the box-shaped the friction profiles. The prefactor $a_1$ corresponds to $\gamma\tb{in}(v\to 0)$, and $a_2$ is the prefactor in the exponential decay, which is roughly unity for all $\varepsilon$. The velocity-scaling for the exponential decay ($a_5\approx 0.5$--$0.6$) is also almost a constant. At high velocities, the velocity-scaling $a_4\approx0.6$ does not depend on $\varepsilon$ \label{tab:gamma_fit} }
    \begin{tabular}{l | c c c c c}
    $\beta\varepsilon$ &$a_1$ &$a_2$&$a_3$&$a_4$&$a_5$\\\hline
    $0.1$ & \ ~9.2  &  0.90  &  2.31 &  0.60  &  0.51 \\
    $1.0$ & 15.4  &  0.96  &  3.52  &  0.59  &  0.61 \\
    $2.0$ & 29.3   & 1.16  &  3.84  &  0.59  &  0.60 \\
    \end{tabular}   
\end{table}
The resulting fit parameters are summarized in \cref{tab:gamma_fit}. $a_1$ corresponds to $\gamma(\vpull\to0)$.
The high-velocity scaling exponent $a_4\approx0.6$ is independent of the interaction strength, indicating that this regime is indeed governed by steric exclusion. Comparing to granular media or fluid dynamics,\cite{Takehara2014Apr,Takehara2010Dec,Gutierrez-Varela2021Sep,Buchholtz1998May} a linear friction scaling with the velocity ($\gamma(v)\propto v$) is expected, while the prefactor is further proportional to the system's density and the tracer size. However, sublinear scaling (as in the present work) has also been reported for at least one order of magnitude of the pulling velocity before converging to the linear scaling at extremely high velocities.\cite{Buchholtz1998May} The reported sublinear scaling coincides with the existence of a gap between the dragged particle and agglomerated bath particles in front. In our polymer membrane, the energy the tracer transfers to one membrane particle during a collision can be passed on to bonded neighbours, possibly pushing particles along the future tracer path. This motion facilitates gap formation and enhanced permeation, resulting in the observed sublinear scaling.
While the remaining fit parameters do not have a direct physical interpretation, their calculation allows for the efficient calculation of $\gamma\tb{in}(v)$ for  implementation in numerical Langevin simulations. 


  \begin{figure*}
      \centering
      \includegraphics[width=\textwidth]{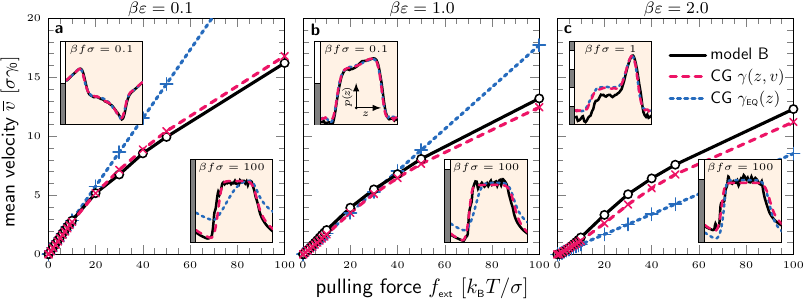}
      \caption{Model B: comparison of microscopic simulations with CG simulation in the presence of a constant external force $f\tb{ext}$ based on Eq.~(\ref{eq:CGLangevin}). The panels depict the tracer's total mean velocity $\overline{v}$ vs. the pulling force, where each panel corresponds to a different interaction strength as indicated above the panels. 
      The black solid line is the full 3D microscopic result, the red dashed line represent 1D Langevin simulation results based on $F\tb{box}$ and $\gamma\tb{box} (v,z)$  (cf. \cref{sec:bipartide}) obtained from targeted MD. The blue dotted lines represent CG simulation results using $F\tb{box}$ and  $\gamma\tb{in}=\gamma\tb{EQ}$. The insets in each panel present examples of the spatial tracer probability $p(z)$ in log-scale for weak (upper left) and strong driving (lower right). The gray-white block pattern on the left axis of the insets reflects orders of magnitude for a rough comparison.   
      \label{fig:compareMDCG}}
  \end{figure*}

\subsubsection{Equilibrium friction}
\begin{table}[b]
    \centering
    \caption{Comparison of equilibrium friction coefficients (in units of $\gamma\tb{0}$) inside the membrane according to \cref{eq:MB_gamma_est}  with known values from equilibrium simulations with multiple tracers.\cite{Milster24} For the value of $\gamma\tu{EQ}\tb{in}$, $\gamma\tb{in}(v)$ is based on the fit results from \cref{eq:gamma_fit} in \cref{tab:gamma_fit}. The last two columns display the constrained friction coefficient $\gamma\tu{PMF}\tb{in}(\vpull=0.01\sigma\gamma\tb{0})$ (PMF-based friction), and the corresponding fit values $a_1$, for comparison.}
    \begin{tabular}{c | r r | r r }
    & \multicolumn{2}{c |}{equilibrium} &  \multicolumn{2}{c}{$\vpull\to 0 $} \\
        $\beta\varepsilon$ \, &  \, $\gamma\tb{in}\tu{EQ}$ & \,\, from \cite{Milster24} & $ \,\gamma\tb{in}\tu{PMF} $ & ~~~~~$a_1$~~\\\hline
        $0.1$ \,  &   6.5~&   6.0~~~~&     8.6~~&    9.2~\\
        $1.0$ \,  &  10.5~&   9.4~~~~&    15.8~~&   15.4~\\ 
        $2.0$ \,  &  15.8~&  19.2~~~~&    27.6~~&   29.3~\\
    \end{tabular}
    
    \label{tab:gamma_reweigh}
\end{table}

Following Onsager's regression hypothesis, our velocity-dependent friction $\gamma\tu{fit}\tb{in}(v)$ should be related to the equilibrium friction $\gamma\tu{EQ}\tb{in}$ for sufficiently weak driving.  
In an equilibrium scenario, the tracer particle exhibits different velocities with $\del F / \del z=0$ and $f\tb{ext}=0$, while the velocities follow a normal distribution $\left.p\tb{EQ}(v)\right|\tb{in}\propto\exp\left(-\beta m v^2/2\right)$. Hence, we propose to estimate the equilibrium friction within the membrane as a Boltzmann-weighted average given by
\begin{align}
    \gamma\tu{EQ}\tb{in}=\int_{-\infty}^{\infty}\mathrm{d}v\,\left.p\tb{EQ}(v)\right|\tb{in} \gamma\tb{in}(v).
    \label{eq:MB_gamma_est}
\end{align}
The results of the reweighing is given in \cref{tab:gamma_reweigh}. For the repulsive ($\beta\varepsilon=0.1$) and the weakly interacting ($\beta\varepsilon=1.0$) membrane, the estimate by \cref{eq:MB_gamma_est} is reasonably close to the equilibrium value as well as the one derived in earlier for a system with multiple tracer particles.\cite{Milster24} In the case of the strongly interacting membrane ($\beta\varepsilon=2.0$), we observe an underestimated friction, indicating that the local polymer dynamics around the tracer slightly differ from the equilibrium scenario. 

In summary, calculating an equilibrium friction coefficient $\gamma\tu{EQ}\tb{in}$ for a system without external driving with data from driven simulations is indeed possible via Eqs.~(\ref{eq:gamma_fit}) and (\ref{eq:MB_gamma_est}).

\subsubsection{Microscopic and CG simulation comparison with constant force-pulling}

So far, we have used constant velocity constraint simulations to coarse-grain the dynamics of our test systems into $F(z)$ and $\gamma(z,v)$ profiles. Now, we want to use these corase-grained profiles to predict the timescales of nonequilibrium tracer transport through a system of interest. To this end, we compare simulations of the full microscopic 3D model B in the presence of a constant driving force that mimics external driving by, e.g., an electrostatic potential, to 1D Langevin (CG) simulations based on \cref{eq:CGLangevin} including the same driving force. We make use of the boxed PMF $F\tb{box}(z)$ and friction $\gamma\tb{box}(z,v)$ with the position-dependent parameters from \cref{tab:meanfieldfit}, while the velocity dependence, $\gamma\tb{in}(v)$, is given by the heuristic \cref{eq:gamma_fit} with parameters provided in \cref{tab:gamma_fit}. To investigate the impact of an explicit consideration of the friction velocity dependence, we performed an additional set of CG simulations where we fixed the friction inside the membrane to the known equilibrium friction $\gamma_{\rm EQ}$,\cite{Milster24} referred to as CG-EQ in the following. 

\cref{fig:compareMDCG} displays the overall mean velocity $\bar{v}$ of the tracer across the system in the steady state (see Appendix \labelcref{app:seim_model_B} for calculation details). In the low-force regime, all models behave identically and exhibit an apparent linear dependence of $\bar{v}$ on $f_{\rm ext}$. This range corresponds to the dissipative linear-response regime in which the equilibrium friction is sufficient to describe the coarse-grained dynamics. Not note that non-linearities do exist in the $v$-vs.-$f$ curves at small forces ($\beta f\tb{ext} \sigma \lesssim 1$) owing to the spatial setup of the membrane, which have already been analyzed and discussed in great detail in previous work\cite{Kim2022Aug,Milster2023Mar} together with the tracer's spatial distributions $p(z)$. 

In the high-force regime, i.e., at roughly  $\beta f\tb{ext} \sigma \gtrsim 10$, the microscopic and the CG result with $\gamma(z,v)$ deviate from CG-EQ and exhibit the same nonlinear trend. For $\beta\varepsilon=0.1$ and  $\beta\varepsilon=1$  (\cref{fig:compareMDCG}a,b), $v(f)$ becomes sublinear as a result of the diverging friction at high velocities. Interestingly, for the highly attractive membrane ($\beta\varepsilon=2$, \cref{fig:compareMDCG}c), the velocity-dependent friction leads to an increased transport. This is in line with the observation  (see \cref{fig:gamma_vs_v}c) that $\gamma(v)$ exhibits a minimum undershooting the equilibrium value. For all interaction strengths we expect the mean velocity to saturate at even higher forces due to the eventual friction divergence. 

The differences between the models become also clear in the spatial profiles presented in the insets in \cref{fig:compareMDCG}. The spatial profiles have been discussed and explained already in our previous work,\cite{Kim2022Aug,Milster2023Mar} here we only summarize the most important findings. For weak driving and repulsive interactions ($\beta\varepsilon=0.1$) tracer particles accumulate (slow down) before entering, and are depleted (accelerated) inside the membrane. For the attractive interactions ($\beta\varepsilon\in\left\{1.0, 2.0\right\}$), we observe an accumulation (slowing down) of tracers within the molecular network. In the case of the strongly interacting membrane and rather slow pulling ($\beta\varepsilon=2$ and $\vpull/v\tb{0}=1$), tracers are hindered to leave the membrane, leading to a pronounced agglomeration. Strong driving removes all these differences and tracer agglomerate in the membrane, due to the diverging friction.

Concerning the performance of CG models, both CG-EQ and the velocity-dependent friction CG model perform practically identical for weak driving (linear-response regime). Under high driving forces, the velocity-dependent model is in better agreement with the microscopic model B reference data. 
For the strongly interacting membrane, both CG models capture the overall trend and the probability profiles, but deviate from the 3D references with rather slow driving, which is likely due to sampling issues of the 3D model in the low probability regions. For strong driving, the sampling improves and we recognize that the velocity-dependent CG model aligns well with the reference and outperforms CG-EQ.


\section{Conclusion}
We have compared different approaches for coarse-graining the microscopic dynamics of a tracer molecule driven through fluctuating, responsive media in two model systems (A and B) by mapping consistently on the potential of mean force and velocity- and position dependent friction profiles.  In particular, we scrutinized the validity regime of applications of the Jarzynski's equality and its second order cumulant approximation, and compared them to more heuristic as well as domain-informed approaches. Noteworthy, the corresponding PMF is independent of the pulling velocity, and only the friction accounts for the additional velocity-dependent effects. The friction can have a non-monotonic dependence on the pulling velocity, specific to the coupling to the fluctuating environment, i.e., to the relevant interacting degrees of freedom with a certain timescale distribution.

The overall trend for the velocity-dependence of the friction in the polymer membrane (model B) was captured by all our friction estimates, but only if the PMF is known (e.g. from slow-pulling simulations), we obtained reliable friction (PMF-based) values for high velocities. The cumulant approximation overestimates the dissipation at high velocities due to the highly correlated dynamics and polymer deformations, resulting in non-normal distributions of the work and the forces, respectively. Jarzynski's method, in contrast, captures the correct values for small velocities, but converges slowly and erratically for intermediate velocities on similar timescale than the system, resulting in an underestimation of the friction. For an efficient and pragmatic coarse-graining, we approximated the PMF and friction profiles by analytical and heuristic (box) functions accounting for the velocity-dependent dissipation. We emphasize again that the applicability of the domain-informed method requires prior knowledge of the overall shape of both PMF and friction coefficient profile. Employing them in 1D Langevin equation simulations reproduced the same spatial probability profiles and mean velocities as the full 3D microscopic simulations.

To apply our approach to other systems of interest, as noted in the Introduction, we advise one to first obtain the equilibrium PMF of a fast converging method, such as the 2\textsuperscript{nd} order cumulant approximation of the Jarzynski's equality used in dcTMD, to determine the shape of the PMF. The pulling should be sufficiently slow to allow the tracer's surrounding to sample its equilibrium phase space, yet fast enough to generate sufficient trajectories. Once the general shape of the PMF is known, one can use an improved estimator such as our PMF-based approach to determine the velocity-dependent friction at higher velocities as described. 

We note that a more fundamental approach that circumvents parametrization of the friction profile is the consideration of memory kernels (as demonstrated for model B in equilibrium~\cite{Milster24}) and the generation of a suitable colored noise from well-defined projection methods,\cite{Widder22,Netz24} which, however, requires significantly more profound, high-resolution sampling and intricate data analysis, in particular for position-dependent kernels,~\cite{position} and in the presence of external fields.~\cite{Koch24,Netz24} Therefore, our presented approach is fast and efficient in the coarse-graining of the dynamics of functional materials under external driving on a molecular level.

\section*{Supplementary Material}
 In the Supplementary Materials we provide animations illustrating the dynamic tracer--barrier interplay (responsive barrier model A) in the regime of imperfect timescale separation ($\vpull/v\tb{0}=1$, $k/k\tb{0}\in\left\{5, 10, 25\right\}$), i.e., corresponding to \cref{fig:1D_domains}b,e, h. 

\section*{Acknowledgments} The authors would like to thank Dominic Nieder for preliminary insights into the responsive barrier model. This work was supported by the Deutsche Forschungsgemeinschaft (DFG) via the Research Unit FOR 5099 ``Reducing complexity of nonequilibrium systems'' (Project No.~431945604) and Project No.~430195928.  The authors also acknowledge support by the state of Baden-Württemberg through bwHPC and the DFG through grant no INST 39/963-1 FUGG (bwForCluster NEMO) and under Germany's Excellence Strategy - EXC-2193/1 - 390951807 ('LivMatS').

\section*{Author declarations}
\subsection*{Conflict of Interest}
The authors have no conflicts to disclose.
\subsection*{Author Contributions}
{\bfseries Sebastian Milster:} Conceptualization (equal); Data curation (lead);
Formal analysis (lead); Investigation (equal); Visualization (lead);
Writing – review \& editing (equal). {\bfseries Joachim Dzubiella:} Conceptualization (equal); Formal analysis (supporting); Investigation (supporting); Writing – review \& editing (equal).  {\bfseries Gerhard Stock:} Conceptualization (equal); Formal analysis (supporting); Investigation (supporting); Writing – review \& editing (equal). {\bfseries Steffen Wolf:} Conceptualization (equal); Data curation (supporting);
Formal analysis (lead); Investigation (equal); Writing – review \& editing (equal). 
\section*{Data availability}
The data that support the findings of this study are available
from the corresponding author upon reasonable request.

\section*{Appendix}
\appendix

\begin{table}[b]
    \centering
    \caption{Parameter summary of for Model A: responsive-barrier crossing \label{tab:barrier_para}}
    \begin{tabular}{l l l}
    description & symbol & value\\ \hline
         solvent friction & $\gamma\tb{0}$ & unit \\
         barrier width & $\sigma\tb{b}$ & unit \\
         system size & $L$ & $10\,\sigma\tb{b}$ \\
         barrier height & $\beta\eps\tb{b}$ & $3$ \\
         barrier stiffness & $k$ & $\in\left[5, 100\right]$ $ \kT\sigma^{-2}\tb{b}$ \\
         pulling velocity & $\vpull$ & $\in\left[0.1, 10\right]$ $\sigma\tb{b}\gamma\tb{0}$ \\
         time increment & $\mathrm{d}t$ & $0.01\gamma\tb{0}^{-1}$
    \end{tabular}    
       \ \\
    \vspace*{1cm}
    \ \\
    \centering
    \caption{Parameter summary for Model B: polymer membrane permeation  \label{tab:membrane_para} }
    \begin{tabular}{l l l}
    description & symbol & value\\ \hline
         solvent friction & $\gamma\tb{0}$ & unit \\
         tracer and polymer size & $\sigma$ & unit \\
         system size in $z$ & $L_z$ & $42.70\,\sigma$ \\
         system size in $x$ and $y$ & $L_{x}, L_y$ & $12.20\,\sigma$ \\
         cross-linker number & $N\tb{x}$ & $104$ \\
         chain-monomner number & $N\tb{mer}$ & $1920$ \\
         polymer volume fraction & $\phi$ & $0.32$ \\
         pulling velocity & $\vpull$  & $\in\left[0.001, 50\right]$ $\sigma\gamma\tb{0}$ \\
         interaction strength & $\beta\varepsilon$ & $\in\left\{0.1, 1.0, 2.0 \right\}$\\
         time increment & $\mathrm{d}t$ & $0.001\gamma\tb{0}^{-1}$
    \end{tabular}    
    \ \\
    \vspace*{1cm}
    \ \\
 \centering
    \caption{PMF fit results for the assumption of box-shaped profiles.}
    \begin{tabular}{c | c c c}
         $\beta\varepsilon$ & $\beta F\tb{in}$ & $d/\sigma$ &  $q$ \\\hline
         0.1 & 1.92  & 11.2 & 8\\ 
         1.0 & -3.34 & 12.6 & 8\\
         2.0 & -11.7  & 13.4 & 8\\      
    \end{tabular}    
    \label{tab:meanfieldfit}
  
\end{table}
\section{Simulation Details Model A\label{app:seim_model_A}}

The integration was performed using a stochastic Heun-integration scheme,\cite{Garcia-Palacios1998Dec} which we carried out with GNU Octave.\cite{GNUOctave} An ensemble of $N\tb{b}=10000$ independent barrier particles is initially equilibrated for 5000 timesteps. In the targeted simulation, the tracer starts at $z=0$ and is pulled with constant velocity $\vpull$ to $z=L$, where $\vpull$ and the time increment $\mathrm{d}t=0.01\gamma\tb{0}^{-1}$ define the number of timesteps $N\tb{steps}=L/(\vpull\mathrm{d}t)$. The forces were ouput each timestep and the analysis results were subsequently averaged into 200 bins. \Cref{tab:barrier_para} summarizes the model parameters.



\section{Simulation Details Model B\label{app:seim_model_B}}

Simulations with model B were carried out with the software package LAMMPS.\cite{LAMMPS} The force-field and the membrane's network topology is identical with our previous work.\cite{Milster24} The most important specifications and parameters are summarized in \cref{tab:membrane_para}. 

For all simulations a single tracer is initially placed at $z=0$ (and random in $x$ and $y$), and pulled along $z$ towards a previously equilibrated membrane. Every time the tracer crosses the box boundary $z=0=L$, the membrane is replaced by a new equilibrated one. This ensures that possible membrane deformations after one permeation do not affect the subsequent trial. 

Concerning the constant-velocity simulations, the constraint for LAMMPS  explained in \cref{sec:v_constraint} is achieved by initializing the tracer velocity in $z$-direction $v_z=v\tb{pull}$, and fixing the force $f_z=0$ for simulation. Note that the forces from the pair-wise interactions are still computed. In fact these are $f\tb{unbiased}=-f\tb{ext}$, the core data of the analysis, which is preaveraged in time windows ($\Delta t = 0.1 \sigma / v\tb{pull}$) corresponding to spatial averages with bin size $\Delta z\approx 0.1 \sigma$. For the data presented in \cref{fig:model_B_profiles} at the pulling velocities $\vpull/v\tb{0}\in\left\{0.001, 0.05, 1\right\}$ we analyzed approx. 120, 2300, and 3000 statistically independent trajectories, respectively.

Regarding the constant-force simulations, the tracer position and velocity is recorded. For consecutive time windows $k$ ($1000$ timesteps long) LAMMPS calculates the temporary probability density $p_{k}(z)$, and the corresponding local time-averaged velocity $\langle v_k(z)\rangle_{t}$. Each window probability is normalized, i.e.,  $\int\mathrm{d}z~p_{k}(z)=1$, and we find the steady-state probability density
$p(z)=\frac{1}{N_k}\sum_k^{N_k}p_k(z)$, and the local ensemble average of the (steady-state) velocity $\langle v(z)\rangle=\frac{1}{N_k}\sum_k^{N_k}p_k(z)\langle v_k(z)\rangle_t$. The overall velocity is computed via $\bar{v}=\int\mathrm{d}z~ p(z) \langle v(z)\rangle$, which is identical with $\bar{v}=\langle L /t\tb{end} \rangle$, where $t\tb{end}$ corresponds to the time one tracer needs to travel the system
length $L$.


\section{Box-shaped profile fits and the PMF-based friction \label{app:MFfit}}
\begin{figure}[t]
    \centering
    \includegraphics[width=1\linewidth]{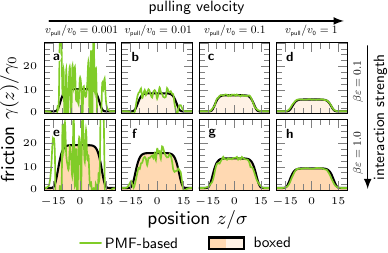}
    \caption{PMF-based friction profiles and boxed friction profile fits for different pulling velocities (see labels above panels) and for different interaction strengths (see labels on the r.h.s.).}
    \label{fig:macrofriction_profile}
\end{figure}

We used the force profiles ($\langle f\tb{ext}\rangle$) from the rather slow pulling simulations ($\vpull\in\left\{0.001, 0.1\right\}$) to first fit a unique $F\tb{in}$, as well as $q$, and $d$, with $z\tb{com}=L/2$ (cf. \cref{sec:bipartide}). The fit results are summarized in \cref{tab:meanfieldfit}. We fixed these values to further fit all $\gamma\tb{in}(v)$, which are practically identical with the PMF-based friction averaged in the central membrane domain. The position- and velocity-dependent friction profiles [computed via \cref{eq:macroscopic_friction}] are displayed in \cref{fig:macrofriction_profile}.

The friction results are presented in \cref{fig:gamma_vs_v}.
   %

\section{Integration of 1D Langevin equation with position- and velocity-dependent friction}\label{app:kinetic_integration}

Given that $F(z)$ and $\gamma(z,v)$ are known, the coarse-grained dynamics described by \cref{eq:CGLangevin} can be efficiently simulated.\cite{Hutter1998} Due to the multiplicative nature of the noise term, i.e., the friction $\gamma(z,v)$ is position- and velocity-dependent, the resulting distribution in space and velocity domain depend on interpretation of the stochastic integral.\cite{Sokolov2010Oct} In the kinetic interpretation (also referred to as the Hänggi-Klimontovich or thermodynamic interpretation \cite{Hanggi1982Aug,Klimontovich1994Aug}) the correct thermodynamic distributions are retained and no additional terms in the Langevin equation are needed to compensate for the \emph{spurious drift}.\cite{Risken1996} Instead, it is already accounted for in the integration scheme.


Although \cref{eq:CGLangevin} is in fact an underdamped description of the position, one can interpret it as an overdamped Langevin equation in the velocity domain with \emph{velocity drift} $A_v$, and \emph{velocity diffusion} $D_v$, reading
 \begin{align}
     A_v(z,v)&=-\frac{1}{m}\frac{\mathrm{d}F(z)}{\mathrm{d}z}-\gamma(z,v)v+\frac{1}{m}f\tb{ext},\\
     D_v(z,v)&=\frac{\kT}{m}\gamma(z,v).
 \end{align}
The drift term can be written as
\begin{align}
    A_v(z,v)=-\beta  D_v(z,v) \delfrac{E(v)|_z}{v},
\end{align}
and the integration scheme will fullfill $p(v)|_z\propto\exp(-\beta E(v)|_z)$. 
\ \\

Following the scheme,\cite{Hutter1998} the predictor is an Euler–Maruyama step
 \begin{align}
     z\tb{p}&=z+v\mathrm{d}t, \\
     v\tb{p}&=v + A_v(z,v)\mathrm{d}t+\sqrt{2D_v(z,v)}\,\mathrm{d}\mathcal{W}_t,
 \end{align}
 which is corrected in the final step as
  \begin{align}
      z \gets z&+\frac{1}{2}\left( v + v\tb{p}\right)\mathrm{d}t\\
      v \gets v&+\frac{1}{2}\left[A_v(z,v)+A_v(z\tb{p},v\tb{p}) \right]\mathrm{d}t+ \\
      &+\frac{1}{2}\left[\frac{D_v(z\tb{p},v\tb{p})}{D_v(z,v)}+1\right] \sqrt{2D_v(z,v)}\,\mathrm{d}\mathcal{W}_t,
      \end{align}
where  $\mathrm{d}\mathcal{W}_t$ is the standard  Wiener process with $\langle\mathrm{d}\mathcal{W}_t\rangle=0$, and $\langle\mathrm{d}\mathcal{W}_t^2\rangle=\mathrm{d}t$. Note that the random number drawn for $\mathrm{d}\mathcal{W}_t$ is identical in the predictor and the final step.

The simulation was carried out with GNU Octave.\cite{GNUOctave}


\bibliographystyle{./aip+title}
\bibliography{./bib}

\begin{thebibliography}{10}

\bibitem{Keener1998Oct}
J.~Keener and J.~Sneyd,
\newblock {\em {Mathematical Physiology (Interdisciplinary Applied Mathematics,
  8)}},
\newblock Springer, New York, NY, USA, 1998.

\bibitem{Copeland2006}
R.~A. Copeland, D.~L. Pompliano, and T.~D. Meek,
\newblock {Drug{\textendash}target residence time and its implications for lead
  optimization},
\newblock Nat. Rev. Drug Discov. {\bf 5}, 730 (2006).

\bibitem{Pan2013}
A.~C. Pan, D.~W. Borhani, R.~O. Dror, and D.~E. Shaw,
\newblock {Molecular determinants of drug{\textendash}receptor binding
  kinetics},
\newblock Drug Discov. Today {\bf 18}, 667 (2013).

\bibitem{Geise2010Aug}
G.~M. Geise, H.-S. Lee, D.~J. Miller, B.~D. Freeman, J.~E. McGrath, and D.~R.
  Paul,
\newblock {Water purification by membranes: The role of polymer science},
\newblock J. Polym. Sci., Part B: Polym. Phys. {\bf 48}, 1685 (2010).

\bibitem{Agboola2021Feb}
O.~Agboola et~al.,
\newblock {A Review on Polymer Nanocomposites and Their Effective Applications
  in Membranes and Adsorbents for Water Treatment and Gas Separation},
\newblock Membranes {\bf 11}, 139 (2021).

\bibitem{Falk2015Apr}
K.~Falk, B.~Coasne, R.~Pellenq, F.-J. Ulm, and L.~Bocquet,
\newblock {Subcontinuum mass transport of condensed hydrocarbons in nanoporous
  media},
\newblock Nat. Commun. {\bf 6}, 1 (2015).

\bibitem{Brohem2011Feb}
C.~A. Brohem, L.~B. da~Silva~Cardeal, M.~Tiago, M.~S. Soengas, S.~B.
  de~Moraes~Barros, and S.~S. Maria-Engler,
\newblock {Artificial skin in perspective: concepts and applications},
\newblock Pigment Cell Melanoma Res. {\bf 24}, 35 (2011).

\bibitem{Wichterle1960Jan}
O.~Wichterle and D.~L{\ifmmode\acute{\imath}\else\'{\i}\fi}m,
\newblock {Hydrophilic Gels for Biological Use},
\newblock Nature {\bf 185}, 117 (1960).

\bibitem{Moncho-Jorda2020Nov}
A.~Moncho-Jord{\ifmmode\acute{a}\else\'{a}\fi}, A.~B.
  J{\ifmmode\acute{o}\else\'{o}\fi}dar-Reyes,
  M.~Kandu{\ifmmode\check{c}\else\v{c}\fi},
  A.~Germ{\ifmmode\acute{a}\else\'{a}\fi}n-Bellod, J.~M.
  L{\ifmmode\acute{o}\else\'{o}\fi}pez-Romero,
  R.~Contreras-C{\ifmmode\acute{a}\else\'{a}\fi}ceres, F.~Sarabia,
  M.~Garc{\ifmmode\acute{\imath}\else\'{\i}\fi}a-Castro, H.~A.
  P{\ifmmode\acute{e}\else\'{e}\fi}rez-Ram{\ifmmode\acute{\imath}\else\'{\i}\fi}rez,
  and G.~Odriozola,
\newblock {Scaling Laws in the Diffusive Release of Neutral Cargo from Hollow
  Hydrogel Nanoparticles: Paclitaxel-Loaded Poly(4-vinylpyridine)},
\newblock ACS Nano {\bf 14}, 15227 (2020).

\bibitem{Stamatialis2008Feb}
D.~F. Stamatialis, B.~J. Papenburg, M.~Giron{\ifmmode\acute{e}\else\'{e}\fi}s,
  S.~Saiful, S.~N.~M. Bettahalli, S.~Schmitmeier, and M.~Wessling,
\newblock {Medical applications of membranes: Drug delivery, artificial organs
  and tissue engineering},
\newblock J. Membr. Sci. {\bf 308}, 1 (2008).

\bibitem{Hoffman1987Dec}
A.~S. Hoffman,
\newblock {Applications of thermally reversible polymers and hydrogels in
  therapeutics and diagnostics},
\newblock J. Controlled Release {\bf 6}, 297 (1987).

\bibitem{Lu2011Jun}
Y.~Lu and M.~Ballauff,
\newblock {Thermosensitive core{\textendash}shell microgels: From colloidal
  model systems to nanoreactors},
\newblock Prog. Polym. Sci. {\bf 36}, 767 (2011).

\bibitem{Angioletti-Uberti2015Jul}
S.~Angioletti-Uberti, Y.~Lu, M.~Ballauff, and J.~Dzubiella,
\newblock {Theory of Solvation-Controlled Reactions in Stimuli-Responsive
  Nanoreactors},
\newblock J. Phys. Chem. C {\bf 119}, 15723 (2015).

\bibitem{Roa2017Sep}
R.~Roa, W.~K. Kim, M.~Kandu{\ifmmode\check{c}\else\v{c}\fi}, J.~Dzubiella, and
  S.~Angioletti-Uberti,
\newblock {Catalyzed Bimolecular Reactions in Responsive Nanoreactors},
\newblock ACS Catal. {\bf 7}, 5604 (2017).

\bibitem{diVentra08}
M.~{Di Ventra},
\newblock {\em Electrical transport in nanoscale systems},
\newblock Cambridge University Press, Cambridge, 2008.

\bibitem{Leitner09}
D.~M. Leitner and J.~E. Straub,
\newblock {\em Proteins: Energy, Heat and Signal Flow},
\newblock Taylor and Francis/CRC Press, London, 2009.

\bibitem{Loos2024Feb}
S.~A.~M. Loos,
\newblock {Smooth Control of Active Matter},
\newblock Physics {\bf 17} (2024).

\bibitem{Liese2017Jan}
S.~Liese, M.~Gensler, S.~Krysiak, R.~Schwarzl, A.~Achazi, B.~Paulus, T.~Hugel,
  J.~P. Rabe, and R.~R. Netz,
\newblock {Hydration Effects Turn a Highly Stretched Polymer from an Entropic
  into an Energetic Spring},
\newblock ACS Nano {\bf 11}, 702 (2017).

\bibitem{Best10}
R.~B. Best and G.~Hummer,
\newblock Coordinate-dependent diffusion in protein folding,
\newblock Proc. Nat. Acad. Sci. USA {\bf 107}, 1088  (2010).

\bibitem{Schulz12}
J.~C.~F. Schulz, L.~Schmidt, R.~B. Best, J.~Dzubiella, and R.~R. Netz,
\newblock Peptide chain dynamics in light and heavy water: Zooming in on
  internal friction,
\newblock J. Aam. Chem. Soc. {\bf 134}, 6273 (2012).

\bibitem{Hu16}
X.~Hu, L.~Hong, M.~Dean~Smith, T.~Neusius, X.~Cheng, and J.~C. Smith,
\newblock {The dynamics of single protein molecules is non-equilibrium and
  self-similar over thirteen decades in time},
\newblock Nat. Phys. {\bf 12}, 171 (2016).

\bibitem{Rico19}
F.~Rico, A.~Russek, L.~{Gonz{\'a}lez}, H.~{Grubm{\"u}ller}, and S.~Scheuring,
\newblock Heterogeneous and rate-dependent streptavidin{\textendash}biotin
  unbinding revealed by high-speed force spectroscopy and atomistic
  simulations,
\newblock Proc. Nat. Acad. Sci. USA {\bf 116}, 6594 (2019).

\bibitem{Berneche01}
B.~S and R.~B,
\newblock {Energetics of ion conduction through the K+ channel},
\newblock Nature {\bf 414} (2001).

\bibitem{Dzubiella2005Jun}
J.~Dzubiella and J.-P. Hansen,
\newblock {Electric-field-controlled water and ion permeation of a hydrophobic
  nanopore},
\newblock J. Chem. Phys. {\bf 122}, 234706 (2005).

\bibitem{Maffeo2012Dec}
C.~Maffeo, S.~Bhattacharya, J.~Yoo, D.~Wells, and A.~Aksimentiev,
\newblock {Modeling and Simulation of Ion Channels},
\newblock Chem. Rev. {\bf 112}, 6250 (2012).

\bibitem{Kopfer14}
D.~A. K{\"o}pfer, C.~Song, T.~Gruene, G.~M. Sheldrick, U.~Zachariae, and B.~L.
  de~Groot,
\newblock {Ion permeation in K+ channels occurs by direct Coulomb knock-on},
\newblock Science {\bf 346}, 352 (2014).

\bibitem{Jaeger22}
M.~J{\"a}ger, T.~Koslowski, and S.~Wolf,
\newblock {Predicting Ion Channel Conductance via Dissipation-Corrected
  Targeted Molecular Dynamics and Langevin Equation Simulations},
\newblock J. Chem. Theory Comput. {\bf 18}, 494 (2022).

\bibitem{Mitchell61}
P.~Mitchell,
\newblock {Coupling of Phosphorylation to Electron and Hydrogen Transfer by a
  Chemi-Osmotic type of Mechanism},
\newblock Nature {\bf 191}, 144 (1961).

\bibitem{Zhang2012May}
H.~Zhang and P.~K. Shen,
\newblock {Recent Development of Polymer Electrolyte Membranes for Fuel Cells},
\newblock Chem. Rev. {\bf 112}, 2780 (2012).

\bibitem{Kim2022AugActive}
Y.~Kim, S.~Joo, W.~K. Kim, and J.-H. Jeon,
\newblock {Active Diffusion of Self-Propelled Particles in Flexible Polymer
  Networks},
\newblock Macromolecules {\bf 55}, 7136 (2022).

\bibitem{Zwanzig01}
R.~Zwanzig,
\newblock {\em Nonequilibrium Statistical Mechanics},
\newblock Oxford University, Oxford, 2001.

\bibitem{Mori65}
H.~Mori,
\newblock Transport, collective motion, and {Brownian} motion,
\newblock Progress of theoretical physics {\bf 33}, 423 (1965).

\bibitem{Grabert82}
H.~Grabert,
\newblock {\em Projection operator techniques in nonequilibrium statistical
  mechanics},
\newblock Springer, Berlin, 1982.

\bibitem{Wolf20}
S.~Wolf, B.~Lickert, S.~Bray, and G.~Stock,
\newblock Multisecond ligand dissociation dynamics from atomistic simulations,
\newblock Nat. Commun. {\bf 11}, 2918 (2020).

\bibitem{Straub87}
J.~E. Straub, M.~Borkovec, and B.~J. Berne,
\newblock Calculation of dynamic friction on intramolecular degrees of freedom,
\newblock J. Phys. Chem. {\bf 91}, 4995  (1987).

\bibitem{Singer1994Sep}
I.~L. Singer,
\newblock {Friction and energy dissipation at the atomic scale: A review},
\newblock J. Vac. Sci. Technol., A {\bf 12}, 2605 (1994).

\bibitem{Setny13}
P.~Setny, R.~Baron, P.~M. Kekenes-Huskey, J.~A. McCammon, and J.~Dzubiella,
\newblock {Solvent fluctuations in hydrophobic cavity-ligand binding kinetics},
\newblock Proc. Natl. Acad. Sci. USA {\bf 110}, 1197 (2013).

\bibitem{Post22}
M.~Post, S.~Wolf, and G.~Stock,
\newblock Molecular origin of driving-dependent friction in fluids,
\newblock J. Chem. Theory Comput. {\bf 18}, 2816 (2022).

\bibitem{Dalton24}
B.~A. Dalton, H.~Kiefer, and R.~R. Netz,
\newblock {The role of memory-dependent friction and solvent viscosity in
  isomerization kinetics in viscogenic media},
\newblock Nat. Commun. {\bf 15}, 1 (2024).

\bibitem{Milster24}
S.~Milster, F.~Koch, C.~Widder, and T.~Schilling,
\newblock {Tracer dynamics in polymer networks: Generalized Langevin
  description},
\newblock J. Chem. Phys. {\bf 160}, 094901 (2024).

\bibitem{Chipot}
A.~P.~e. Christophe Chipot~(editor),
\newblock {\em Free energy calculations},
\newblock Springer series in chemical physics, volume 86, Springer, 1 edition,
  2007.

\bibitem{Risken1996}
H.~Risken,
\newblock {\em {The Fokker-Planck Equation}},
\newblock Springer, Berlin, Germany, 1996.

\bibitem{Oettinger05}
{H. C. \"Ottinger},
\newblock {\em Beyond Equilibrium Thermodynamics},
\newblock Wiley, Hoboken, 2005.

\bibitem{Haenggi97}
P.~H{\"a}nggi,
\newblock Generalized {Langevin} equations: A useful tool for the perplexed
  modeller of nonequilibrium fluctuations?,
\newblock in {\em Stochastic Dynamics}, edited by L.~Schimansky-Geier and
  T.~P{\"o}schel, pages 15--22, Berlin, Heidelberg, 1997, Springer Berlin
  Heidelberg.

\bibitem{Fortini2014Oct}
A.~Fortini, D.~de~las Heras, J.~M. Brader, and M.~Schmidt,
\newblock {Superadiabatic Forces in Brownian Many-Body Dynamics},
\newblock Phys. Rev. Lett. {\bf 113}, 167801 (2014).

\bibitem{Shinjo1993Mar}
K.~Shinjo and M.~Hirano,
\newblock {Dynamics of friction: superlubric state},
\newblock Surf. Sci. {\bf 283}, 473 (1993).

\bibitem{Erdmann2000May}
U.~Erdmann, W.~Ebeling, L.~Schimansky-Geier, and F.~Schweitzer,
\newblock {Brownian particles far from equilibrium},
\newblock Eur. Phys. J. B {\bf 15}, 105 (2000).

\bibitem{Buchholtz1998May}
V.~Buchholtz and T.~P{\ifmmode\ddot{o}\else\"{o}\fi}schel,
\newblock {Interaction of a granular stream with an obstacle},
\newblock Granular Matter {\bf 1}, 33 (1998).

\bibitem{Gutierrez-Varela2021Sep}
O.~Guti{\ifmmode\acute{e}\else\'{e}\fi}rrez-Varela and R.~Santamaria,
\newblock {Molecular nature of the drag force},
\newblock J. Mol. Liq. {\bf 338}, 116466 (2021).

\bibitem{Takehara2010Dec}
Y.~Takehara, S.~Fujimoto, and K.~Okumura,
\newblock {High-velocity drag friction in dense granular media},
\newblock Europhys. Lett. {\bf 92}, 44003 (2010).

\bibitem{Zhu1990Sep}
S.-B. Zhu,
\newblock {Velocity distributions in nonlinear systems},
\newblock Phys. Rev. A {\bf 42}, 3374 (1990).

\bibitem{Takehara2014Apr}
Y.~Takehara and K.~Okumura,
\newblock {High-Velocity Drag Friction in Granular Media near the Jamming
  Point},
\newblock Phys. Rev. Lett. {\bf 112}, 148001 (2014).

\bibitem{Voigtmann2013Nov}
{\relax Th}.~Voigtmann and M.~Fuchs,
\newblock {Force-driven micro-rheology},
\newblock Eur. Phys. J. Spec. Top. {\bf 222}, 2819 (2013).

\bibitem{Fusco2005Jan}
C.~Fusco and A.~Fasolino,
\newblock {Velocity dependence of atomic-scale friction: A comparative study of
  the one- and two-dimensional Tomlinson model},
\newblock Phys. Rev. B {\bf 71}, 045413 (2005).

\bibitem{Loos2024May}
S.~A.~M. Loos, S.~Monter, F.~Ginot, and C.~Bechinger,
\newblock {Universal Symmetry of Optimal Control at the Microscale},
\newblock Phys. Rev. X {\bf 14}, 021032 (2024).

\bibitem{Matsukawa1994Jun}
H.~Matsukawa and H.~Fukuyama,
\newblock {Theoretical study of friction: One-dimensional clean surfaces},
\newblock Phys. Rev. B {\bf 49}, 17286 (1994).

\bibitem{Meyer19}
H.~Meyer, T.~Voigtmann, and T.~Schilling,
\newblock On the dynamics of reaction coordinates in classical, time-dependent,
  many-body processes,
\newblock J. Comput. Phys. {\bf 150}, 174118 (2019).

\bibitem{Meyer20}
H.~Meyer, S.~Wolf, G.~Stock, and T.~Schilling,
\newblock A numerical procedure to evaluate memory effects in non-equilibrium
  coarse-grained models,
\newblock Adv. Theory Simul. {\bf 111}, 2000197 (2020).

\bibitem{Schmid2023Feb}
F.~Schmid,
\newblock {Understanding and Modeling Polymers: The Challenge of Multiple
  Scales},
\newblock ACS Polym. Au {\bf 3}, 28 (2023).

\bibitem{Ariskina2024Sep}
K.~Ariskina, G.~Galli{\ifmmode\acute{e}\else\'{e}\fi}ro, and A.~Obliger,
\newblock {Confined fluid dynamics in a viscoelastic, amorphous, and
  microporous medium: Study of a kerogen by molecular simulations and the
  generalized Langevin equation},
\newblock J. Chem. Phys. {\bf 161}, 124901 (2024).

\bibitem{Klippenstein2021May}
V.~Klippenstein, M.~Tripathy, G.~Jung, F.~Schmid, and N.~F.~A. van~der Vegt,
\newblock {Introducing Memory in Coarse-Grained Molecular Simulations},
\newblock J. Phys. Chem. B {\bf 125}, 4931 (2021).

\bibitem{Dalton2024Oct}
B.~A. Dalton and R.~R. Netz,
\newblock {pH Modulates Friction Memory Effects in Protein Folding},
\newblock Phys. Rev. Lett. {\bf 133}, 188401 (2024).

\bibitem{Straube2020Jul}
A.~V. Straube, B.~G. Kowalik, R.~R. Netz, and
  F.~H{\ifmmode\ddot{o}\else\"{o}\fi}fling,
\newblock {Rapid onset of molecular friction in liquids bridging between the
  atomistic and hydrodynamic pictures},
\newblock Commun. Phys. {\bf 3}, 1 (2020).

\bibitem{Jung2017Jun}
G.~Jung, M.~Hanke, and F.~Schmid,
\newblock {Iterative Reconstruction of Memory Kernels},
\newblock J. Chem. Theory Comput. {\bf 13}, 2481 (2017).

\bibitem{Wolf18}
S.~Wolf and G.~Stock,
\newblock Targeted molecular dynamics calculations of free energy profiles
  using a nonequilibrium friction correction,
\newblock J. Chem. Theory Comput. {\bf 14}, 6175  (2018).

\bibitem{Schilling2022Aug}
T.~Schilling,
\newblock {Coarse-grained modelling out of equilibrium},
\newblock Phys. Rep. {\bf 972}, 1 (2022).

\bibitem{Koch24}
F.~Koch, J.~Erle, and T.~Schilling,
\newblock {Nonequilibrium solvent response force: What happens if you push a
  Brownian particle},
\newblock Phys. Rev. Res. {\bf 6}, L012032 (2024).

\bibitem{Kosztin2006Feb}
I.~Kosztin, B.~Barz, and L.~Janosi,
\newblock {Calculating potentials of mean force and diffusion coefficients from
  nonequilibrium processes without Jarzynski's equality},
\newblock J. Chem. Phys. {\bf 124}, 64106. (2006).

\bibitem{Kubo1966Jan}
R.~Kubo,
\newblock {The fluctuation-dissipation theorem},
\newblock Rep. Prog. Phys. {\bf 29}, 255 (1966).

\bibitem{BarratHansen}
J.-L. Barrat and J.-P. Hansen,
\newblock {\em Basic Concepts for Simple and Complex Liquids},
\newblock Cambridge University Press, 2003.

\bibitem{Dudko06}
O.~K. Dudko, G.~Hummer, and A.~Szabo,
\newblock {Intrinsic Rates and Activation Free Energies from Single-Molecule
  Pulling Experiments},
\newblock Phys. Rev. Lett. {\bf 96}, 108101 (2006).

\bibitem{Jarzynski97}
C.~Jarzynski,
\newblock Nonequilibrium equality for free energy differences,
\newblock Phys. Rev. Lett. {\bf 78}, 2690 (1997).

\bibitem{Wolf23}
S.~Wolf, M.~Post, and G.~Stock,
\newblock Path separation of dissipation-corrected targeted molecular dynamics
  simulations of protein-ligand unbinding,
\newblock J. Comput. Phys. {\bf 158}, 124106 (2023).

\bibitem{Gazuz2013Mar}
I.~Gazuz and M.~Fuchs,
\newblock {Nonlinear microrheology of dense colloidal suspensions: A
  mode-coupling theory},
\newblock Phys. Rev. E {\bf 87}, 032304 (2013).

\bibitem{Gnann2011Feb}
M.~V. Gnann, I.~Gazuz, A.~M. Puertas, M.~Fuchs, and {\relax Th}.~Voigtmann,
\newblock {Schematic models for active nonlinear microrheology},
\newblock Soft Matter {\bf 7}, 1390 (2011).

\bibitem{Gazuz2009Jun}
I.~Gazuz, A.~M. Puertas, {\relax Th}.~Voigtmann, and M.~Fuchs,
\newblock {Active and Nonlinear Microrheology in Dense Colloidal Suspensions},
\newblock Phys. Rev. Lett. {\bf 102}, 248302 (2009).

\bibitem{Yasuda1969}
H.~Yasuda, A.~Peterlin, C.~K. Colton, K.~A. Smith, and E.~W. Merrill,
\newblock Permeability of solutes through hydrated polymer membranes - part
  iii. theoretical background for the selectivity of dialysis membranes,
\newblock Die Makromol. Chemie {\bf 126}, 177 (1969).

\bibitem{Amsden1998}
B.~Amsden,
\newblock Solute diffusion within hydrogels. mechanisms and models,
\newblock Macromolecules {\bf 31}, 8382 (1998).

\bibitem{Kim2020}
W.~K. Kim, R.~Chudoba, S.~Milster, R.~Roa, M.~Kandu\v\{c\}, and J.~Dzubiella,
\newblock Tuning the selective permeability of polydisperse polymer networks,
\newblock Soft Matter {\bf 16}, 1 (2020).

\bibitem{Prandtl1928Jan}
L.~Prandtl,
\newblock {Ein Gedankenmodell zur kinetischen Theorie der festen
  K{\ifmmode\ddot{o}\else\"{o}\fi}rper},
\newblock Z. angew. Math. Mech. {\bf 8}, 85 (1928).

\bibitem{Tomlison1929Jun}
G.~A.~T. B.~Sc.,
\newblock {CVI. A molecular theory of friction},
\newblock London, Edinburgh, and Dublin Philosophical Magazine and Journal of
  Science  (1929).

\bibitem{Dong2011Dec}
Y.~Dong, A.~Vadakkepatt, and A.~Martini,
\newblock {Analytical Models for Atomic Friction},
\newblock Tribol. Lett. {\bf 44}, 367 (2011).

\bibitem{Muser2011Sep}
M.~H. M{\ifmmode\ddot{u}\else\"{u}\fi}ser,
\newblock {Velocity dependence of kinetic friction in the Prandtl-Tomlinson
  model},
\newblock Phys. Rev. B {\bf 84}, 125419 (2011).

\bibitem{Gnecco2012Jul}
E.~Gnecco, R.~Roth, and A.~Baratoff,
\newblock {Analytical expressions for the kinetic friction in the
  Prandtl-Tomlinson model},
\newblock Phys. Rev. B {\bf 86}, 035443 (2012).

\bibitem{Schmitt2006May}
C.~Schmitt, B.~Dybiec, P.~H{\ifmmode\ddot{a}\else\"{a}\fi}nggi, and
  C.~Bechinger,
\newblock {Stochastic resonance vs. resonant activation},
\newblock Europhys. Lett. {\bf 74}, 937 (2006).

\bibitem{Doering1992Oct}
C.~R. Doering and J.~C. Gadoua,
\newblock {Resonant activation over a fluctuating barrier},
\newblock Phys. Rev. Lett. {\bf 69}, 2318 (1992).

\bibitem{Daldrop17}
J.~O. Daldrop, B.~G. Kowalik, and R.~R. Netz,
\newblock External potential modifies friction of molecular solutes in water,
\newblock Phys. Rev. X {\bf 7}, 041065 (2017).

\bibitem{Kim2022Aug}
W.~K. Kim, S.~Milster, R.~Roa, M.~Kandu{\ifmmode\check{c}\else\v{c}\fi}, and
  J.~Dzubiella,
\newblock {Permeability of Polymer Membranes beyond Linear Response},
\newblock Macromolecules {\bf 55}, 7327 (2022).

\bibitem{Milster2023Mar}
S.~Milster, W.~K. Kim, and J.~Dzubiella,
\newblock {Feedback-controlled solute transport through chemo-responsive
  polymer membranes},
\newblock J. Chem. Phys. {\bf 158} (2023).

\bibitem{Widder22}
C.~Widder, F.~Koch, and T.~Schilling,
\newblock {Generalized Langevin dynamics simulation with non-stationary memory
  kernels: How to make noise},
\newblock J. Chem. Phys. {\bf 157}, 194107 (2022).

\bibitem{Netz24}
R.~R. Netz,
\newblock {Derivation of the nonequilibrium generalized Langevin equation from
  a time-dependent many-body Hamiltonian},
\newblock Phys. Rev. E {\bf 110}, 014123 (2024).

\bibitem{position}
H.~Vroylandt and P.~Monmarché,
\newblock Position-dependent memory kernel in generalized langevin equations:
  Theory and numerical estimation,
\newblock The Journal of Chemical Physics {\bf 156}, 244105 (2022).

\bibitem{Garcia-Palacios1998Dec}
J.~L. Garc{\ifmmode\acute{\imath}\else\'{\i}\fi}a-Palacios and F.~J.
  L{\ifmmode\acute{a}\else\'{a}\fi}zaro,
\newblock {Langevin-dynamics study of the dynamical properties of small
  magnetic particles},
\newblock Phys. Rev. B {\bf 58}, 14937 (1998).

\bibitem{GNUOctave}
J.~W. Eaton, D.~Bateman, S.~Hauberg, and R.~Wehbring,
\newblock {GNU Octave} version 9.2.0: A high-level interactive language for
  numerical computations.,
\newblock (2024).

\bibitem{LAMMPS}
A.~P. Thompson et~al.,
\newblock {LAMMPS} - a flexible simulation tool for particle-based materials
  modeling at the atomic, meso, and continuum scales,
\newblock Comp. Phys. Comm. {\bf 271}, 108171 (2022).

\bibitem{Hutter1998}
M.~H{\ifmmode\ddot{u}\else\"{u}\fi}tter and H.~C.
  {\ifmmode\ddot{O}\else\"{O}\fi}ttinger,
\newblock {Fluctuation-dissipation theorem, kinetic stochastic integral and
  efficient simulations},
\newblock J. Chem. Soc., Faraday Trans. {\bf 94}, 1403 (1998).

\bibitem{Sokolov2010Oct}
I.~M. Sokolov,
\newblock {Ito, Stratonovich, H{\ifmmode\ddot{a}\else\"{a}\fi}nggi and all the
  rest: The thermodynamics of interpretation},
\newblock Chem. Phys. {\bf 375}, 359 (2010).

\bibitem{Hanggi1982Aug}
P.~H{\ifmmode\ddot{a}\else\"{a}\fi}nggi and H.~Thomas,
\newblock {Stochastic processes: Time evolution, symmetries and linear
  response},
\newblock Phys. Rep. {\bf 88}, 207 (1982).

\bibitem{Klimontovich1994Aug}
Y.~L. Klimontovich,
\newblock {Nonlinear Brownian motion},
\newblock Phys.-Usp. {\bf 37}, 737 (1994).

\end{thebibliography}

\end{document}